\documentclass[twocolumn,nofiglist,notablist]{rsauthor}
\usepackage{graphicx}
\usepackage{amsmath}
\usepackage{amssymb}
\usepackage{amsfonts}
\usepackage{amsthm}
\usepackage{endfloat}
\usepackage{setspace}
\usepackage{verbatim}
\usepackage{geometry}
\usepackage{times}
\usepackage{helvet}
\usepackage{courier}
\usepackage{bm}
\usepackage{url}
\usepackage{dcolumn}
\usepackage{multirow}
\usepackage{color}

\jname{Phil. Trans. R. Soc. B}
\begin{document}

\title[Efficiencies of Modules in Connectomes]{From {\it Caenorhabditis elegans} to the Human Connectome: A Specific Modular Organisation Increases Metabolic, Functional, and Developmental Efficiency}

\author[J. S. Kim and Marcus Kaiser]{Jinseop S. Kim$^{1,2}$ and Marcus Kaiser\thanks{* Author for correspondence ({M.Kaiser@ncl.ac.uk}).}$^{3,4,2,*}$ }

\address{$^1$ Department of Brain and Cognitive Sciences, Massachusetts Institute of Technology, Cambridge, USA\\ $^2$ Department of Brain and Cognitive Sciences, Seoul National University, Seoul, South Korea\\ 
$^3$ School of Computing Science, Newcastle University, Claremont Tower, Newcastle upon Tyne NE1 7RU, UK\\
$^4$ Institute of Neuroscience, Newcastle University, Framlington Place, Newcastle upon Tyne NE2 4HH, UK
}

\label{firstpage}

%\section*{Abstract}
\abstract{ % up to 200 words
The connectome, or the entire connectivity of a neural system represented by network, ranges various scales from synaptic connections between individual neurons to fibre tract connections between brain regions. Although the modularity they commonly show has been extensively studied, it is unclear whether connection specificity of such networks can already be fully explained by the modularity alone. To answer this question, we study two networks, the neuronal network of {\it C. elegans} and the fibre tract network of human brains yielded through diffusion spectrum imaging (DSI). We compare them to their respective benchmark networks with varying modularities, which are generated by link swapping to have desired modularity values but otherwise maximally random. We find several network properties that are specific to the neural networks and cannot be fully explained by the modularity alone. First, the clustering coefficient and the characteristic path length of {\it C. elegans} and human connectomes are both higher than those of the benchmark networks with similar modularity. High clustering coefficient indicates efficient local information distribution and high characteristic path length suggests reduced global integration. Second, the total wiring length is smaller than for the alternative configurations with similar modularity. This is due to lower dispersion of connections, which means each neuron in {\it C. elegans} connectome or each region of interest (ROI) in human connectome reaches fewer ganglia or cortical areas, respectively. Third, both neural networks show lower algorithmic entropy compared to the alternative arrangements. This implies that fewer rules are needed to encode for the organisation of neural systems. While the first two findings show that the neural topologies are efficient in information processing, this suggests that they are also efficient from a developmental point of view. Together, these results show that neural systems are organised in such a way to yield efficient features beyond those given by their modularity alone. 
}

\maketitle

\section{Introduction}
In network representation, neural networks at different levels of organisation ranging from connections between individual neurons to connections between brain regions can be described coherently, if the individual neurons or brain regions are substituted by the nodes and the connection between them by the links. Also, the modular organisation found in different levels of neural networks can be exhibited by network modules, where a module is a subset of the nodes having many connections among them and few to the rest of the network \cite{kaiser_tutorial}. 

The first species to show neural networks are Coelenterates such as {\it Cnidaria} \cite{mackie,arendt}. These animals show a diffuse two-dimensional nerve network called a lattice network. In such networks, neighbours are well connected but there are no long-distance connections. For functionally specialised circuits, however, a regular organisation is unsuitable. Starting with the formation of sensory organs and motor units, neurons segregate in modules; e.g. forming ganglia in the roundworm {\it Caenorhabditis elegans} \cite{white}. Forming such modules, ganglia can process one modality with little interference from neurons processing different kinds of information. At one point of growing complexity of organisms, having one module for one modality or function is not sufficient. An example is processing of visual information in primates where the visual module consists of two network components: nodes that form the dorsal pathway for processing object movement and nodes of the ventral pathway for processing object features such as colour and form. These networks where smaller sub-modules are nested within modules are a type of hierarchical network \cite{kaiser_dyn,meunier,krumnack}.

The modularity $Q$ measures how modular a given network is \cite{newman}. The human brain network for the connections between brain regions or ROIs as well as the neuronal network for the connections between neurons show a high modularity compared to randomly connected networks \cite{meunier} and this modularity is preserved from at least 4 to 40 years \cite{Lim2014CerCor}. However, there are numerous ways of constructing modular networks with a given value of modularity. What are  specific to the chosen biological organisations over alternative modular arrangements and what are the advantages of them? In this article, we address these questions on two different levels of organisation: the connections between individual neurons in {\it C. elegans}, the level of the micro-connectome \cite{seung}, and the connections between different human brain regions, the level of the macro-connectome \cite{defelipe}. To investigate the connection specificity of these networks over alternative arrangements, we employ benchmark networks generated by a link swapping process which is controlled by the simulated annealing algorithm. Such rewired networks can serve as control groups, where the number of connections for each node and the modularity of networks are kept constant.

First, at both levels we find that the clustering coefficient, indicating how well information can be distributed locally, and the characteristic path length, indicating how difficult global integration is, are high compared to alternative networks of similar modularity. This shows a balance between the need for communication within local circuits (high neighbourhood connectivity within modules) and the reduction of interference between modules (fewer shortcuts linking different modules). Indeed, brain disorders such as schizophrenia \cite{sritharan} and epilepsy \cite{chavez} can be linked to changes in local and global efficiency. Second, the total wiring length is smaller compared to the alternative networks of similar modularity. The connectivity of the original network and alternative networks are compared through their network of modules, the coarse-grained network obtained when human brain areas are considered as new nodes instead of the ROIs. We find that the formation of fibre bundles, or the fasciculation, is correlated with the reduced total wiring length. A similar behaviour is observed from the network of neurons and the network of ganglia in {\it C. elegans}. To quantify this bundling behaviour, we introduce the novel measure of dispersion indicating how widely individual nodes are connected to different modules of the network. Third, both neural networks show lower algorithmic entropy than their alternative arrangements. As the algorithmic entropy quantifies the amount of information needed to construct an object, this suggests that fewer rules are needed to encode for the organisation of neural networks and the neural systems are efficiently organised from a developmental point of view \cite{roberts}.

\section{Materials and Methods}
\subsection*{Data}
The human brain network used in this article was from \cite{hagmann}. The connectivity was obtained from $5$ individual subjects using the diffusion spectrum imaging (DSI). The DSI is one of the protocols of diffusion magnetic resonance imaging (dMRI), which detects the diffusion pattern of water molecules in the brain to predict the trajectory of fibre tracts. In the DSI, first the brain was partitioned into anatomical areas called the Brodmann areas and then each of them was subdivided into a certain number of ROIs in such a way that each ROI has a similar surface area. The number of brain areas were chosen to be $R=66$ and the number of ROIs resulted in $N=998$. The ROIs were regarded as nodes and the brain areas as modules. Next, the tractography was constructed from the diffusion pattern and a link was assigned between two ROIs that are connected by the predicted fibre tract. The total number of links was $E=17,865$. 

For {\it C. elegans}, a total of $N=279$ neurons and corresponding $E=2,990$ connections were used. These included $1,584$ unidirectional and $1,406$ bidirectional connections. Biologically, they represent $672$ gap junctions, $1,962$ chemical synapses and $376$ connections where both gap junctions and chemical synapses exist between the neuron pairs. As some network measures are defined only for undirected networks, all the unidirectional connections were replaced by bidirectional ones leading to a total of $2,287$ bidirectional links. Three-dimensional neuron coordinates were used as described in \cite{varier}. The information about the $R=10$ ganglia membership for modules was taken from \cite{ay}.

The network of modules was defined as follows. The modules, corresponding to the anatomical areas for brain or the ganglia for {\it C. elegans}, were regarded as nodes in place of the ROIs or neurons. Correspondingly, two modules were assigned with a link between them only if there is at least one link between a pair of nodes each of which is contained by each module (see Supplementary Figure S1). 

\subsection*{Network measures}
All the calculations, including measurement of modularity and simulated annealing procedure (see below), were performed by custom built codes in C programming language and MATLAB (routines are available at {http://www.biological-networks.org/}). The characteristic path length ($L$) was the average number of connections that have to be passed on the shortest paths between all pairs of network nodes. The clustering coefficient ($C$) was the proportion of actually present connections, out of all possible connections, among network nodes directly connected to a node. It was calculated as the average over all individual nodes of the network \cite{watts}. The small-world index was calculated as $\sigma_{\text{sw}}=(C/C_{\text{rand}}) / (L/L_{\text{rand}})$ or equivalently $\sigma_{\text{sw}}=(C/L) / (C_{\text{rand}}/L_{\text{rand}})$, where $C$ and $L$ defined as above were measured from the observed network and $C_{\text{rand}}$ and $L_{\text{rand}}$ were the average values from $100$ Erd\H{o}s-R\'enyi (ER) random network \cite{humphries}. The generation rule for the ER network was as following. Initially $N$ nodes are given without any connection. At each time step, a link is added between a pair of nodes which are selected among the $N$ nodes at random, avoiding multiple times of selection. This step is repeated until the number of links becomes $E$.  The small-worldness $\sigma_{\text{sw}}$  is larger than 1 for small-world networks, equal to 1 if the ratio between $C/C_{\text{rand}}$ and $L/L_{\text{rand}}$ is the same as for random networks (note that absolute $C$ and $L$ might still differ from those of random networks), and smaller than 1 when the clustering coefficient is smaller and/or the characteristic path length is larger than for random networks.  The total wiring length ($W$) was the sum of the Euclidean distance between all connections of a network when the network nodes are provided with spatial locations. 

\subsection*{Modularity and link swapping}
For a network of $N$-nodes, $E$-links and $R$-modules, whose node index, $i$, runs from $1$ to $N$ and the module to which node $i$ belongs, $q_i$, can take value from $1$ to $R$, the modularity was defined as
\begin{equation}
Q=\frac{1}{2E} \sum_{ij} \left[ A_{ij}-\frac{k_i k_j}{2E}\right]\delta(q_i,q_j),
\end{equation}
where $A_{ij}$ is $(i,j)$ element of the adjacency matrix, $k_i$ is the number of connections, or the degree, of node $i$, and $\delta$ is Kronecker's delta function \cite{newman}. It measures what fraction of the links connect two nodes within one module and its deviation from the case when the links are distributed at random. The modularity can be used for finding the modular structure of a given network when it is unknown. In such a setting, an optimal partitioning of the network nodes is searched, which maximises the modularity of the given network. Therefore, the assignments of nodes to modules are varied while the connections of the nodes are fixed. In this study, however, the predefined modules of respective networks, i.e. the anatomical areas of human brain and the ganglia of {\it C. elegans}, were regarded as fixed. Each node already has its intrinsic module membership. Instead, the connections between nodes were varied by link swapping controlled by simulated annealing. 

The link swapping is a process which a pair of links are selected and then two nodes at an arbitrary end of each link are exchanged. Whereas the degree of each node, as well as its distribution for the entire network, is preserved before and after the manipulation, the modularity of the network can be increased, be decreased, or remain the same depending on the sort of the selected pair of links. It is increased if a pair of links are selected in such a way that at least two nodes at the ends of different links lie in one module and swapping is carried out to connect those two nodes. Likewise, the number of intra-module links determines the modularity of rewired network after the swapping (see Supplementary Figure S1).

To alter the modularity of the networks to have desired values, the selection of link pairs for swapping process was controlled by simulated annealing method as follows \cite{guimera}. At each step, the link swapping is attempted and the amount of change in modularity for the attempt, $\Delta Q$, is calculated. The attempt is accepted with probability $1$ if $\Delta Q \ge 0$ or with probability $e^{\Delta Q/T}$ if $\Delta Q<0$, where $T$ is the control parameter or temperature. Otherwise, the attempt is rejected and the swapping is reversed to recover the original connectivity. When $T \to 0$, link swappings are accepted only when the modularity increases and the simulated annealing becomes equivalent to the greedy algorithm for finding the maximum modularity. Originally, the simulated annealing was devised to avoid trapping into local extrema as the greedy algorithm often does, and $T$ is incrementally decreased from a finite value to infinitesimal so that the swapping happens a certain number of times at each $T$ value. The consequent maximum value during the entire time steps is expected to be the global maximum. In a similar manner, to obtain a network with the desired modularity $Q_d$, one can set the problem to minimise $\lvert Q-Q_d \rvert$. 

However, in this study, we employed a simpler method since the minimisation procedure is computationally expensive and the networks from the two different methods are theoretically equivalent to each other. The alternative method took advantage of the fact that the modularity, unless small fluctuations, converges to a single value for a given temperature. After a sufficient number of link swappings are performed, the connection specificity of the original network is lost and the resulting network has desired modularity but otherwise maximally random. Any choice of network snapshot at this state is statistically identical to each other, and the entire set of such networks is the ensemble of networks with the given modularity. In practice, we first performed $800 \times E$ times of link swapping for a given temperature, and then sampled $100$ network snapshots during additional $200 \times E$ of steps. 

\subsection*{Dispersion}
We introduced the novel measure, dispersion $D$, of a network which shows how widely the connections are distributed across different modules. The dispersion of an individual node $i$ was defined as $D_i = R_i / R$ where $R_i$ is the number of different modules to which the node is connected to (brain areas for the human connectome or ganglia for the {\it C. elegans} connectome) and $R$ is the total number of modules ($66$ and $10$, respectively). The maximum dispersion of a node is 1 in the case where the node is connected to at least one node in all other modules of the network. The dispersion of a network is the average dispersion for all nodes: $D = \sum D_i / N$ where $N$ is the number of nodes ($998$ ROIs for the human connectome and $279$ neurons for the {\it C. elegans} connectome). Note that the modules in this study are anatomical units (brain areas or ganglia) and not the modules defined by network analysis module detection algorithms \cite{kaiser_tutorial}. However, alternative definitions for module can also be applied, and the dispersion could serve as a useful measure for future studies. 

\subsection*{Algorithmic entropy}
Algorithmic entropy was used as a measure for the amount of information the networks bear. It was originally introduced as a conceptual measure for any kind of physical or abstract objects, and later a practical way to quantify it was devised \cite{li}. Assume an object saved in a computer storage device. If the object contains regularities, it can be described by a shorter message leading to less storage usage. A compression algorithm is a standard way to detect such regularities and reduce storage usage, and the compressed data size can give an estimate of the amount of information. To apply this to the neural networks, we saved the networks in the format of unweighted adjacency matrices into $N \times N$ {\tt int8} arrays, whose $(i, j)$ element takes value $1$ if nodes $i$ and $j$ have a connection to each other and otherwise $0$. Any configuration of networks with the same number of nodes $N$ has $N^2$ bytes of data size. Then, minimum compression size for each of the adjacency matrix arrays for the original connectomes as well as the rewired network ensembles for different values of $Q$ was found by the simulated annealing method similar to above. The compression was performed by the {\tt gzip} library which uses the Lempel-Ziv coding \cite{ziv}. The compression ratio, the ratio of the compressed data size to the original size of the array in bytes, was measured to indicate the relative amount of information in the networks. As the adjacency matrix of the networks are symmetric and sparse, more efficient data storing strategy could be devised. Although this can change the quantitative values of the compression ratio, it is unlikely that the qualitative trend of the result from the original and alternative networks would change. 

For simulated annealing, the objective measure to minimise was the compression size $Z$ and the variable was node index assignment. Whereas the assignment of node index, i.e. which node becomes the node $i$, is arbitrary, the shape of the adjacency matrix depends on the index assignment and in turn the compression size depends on the shape. As the algorithmic entropy, by definition, aims to measure the upper bound of the amount of information, the node index assignment needs to minimise the size of compressed array. At each time step of the simulated annealing, the node indices were reassigned by exchanging the indices of two nodes $i$ and $j$, which is equivalent to exchanging the $i$-th and $j$-th column and row of the adjacency matrix. Then $\Delta Z$ was measured by comparing the $Z$ values before and after the reassignment, to determine such a reassignment should be kept or reverted with the probability of $1$ when $\Delta Z \le 0$ or with probability of $e^{-\Delta Z/T}$ when $\Delta Z>0$. $T$ was incrementally decreased from a finite value to infinitesimal so that the index reassignment happens a certain number of time steps at each $T$ value. The global minimum $Z$ during the entire time step was recorded.

\section{Results}
To illustrate the connectivity of the neural networks, we calculated the network measures of the human and {\it C. elegans} connectome. Two relevant measures, $L$ and $C$, were compared to those of the ER random networks with the same number of nodes and links (Table \ref{table1}). First, the characteristic path length $L$, related to the global efficiency of reaching other nodes at the global level, shows the average number of connections that need to be crossed to go from one network node to another. Second, the clustering coefficient $C$, related to the local efficiency of reaching nearby nodes, indicates how well neighbours of a node are connected, i.e. what proportion of potential links between neighbours actually exists. Third, the small-world index $\sigma_{\text{sw}}$ indicates to what extent the fraction of two small-world measures, $C/L$, of a network deviates from that of random networks. Finally, we observed the total wiring length that is the sum of the approximated metric lengths of all individual connections. Note that the Euclidean distance in three dimensions gives an estimate or lower bound of the length of a connection, as the curvature in actual wiring between nodes makes the real distance longer. More information on network measures can be found in \cite{kaiser_tutorial, rubinov}.

The human macro-connectome consists of $R=66$ brain areas (modules), $N=998$ ROIs (nodes), and $E=17,865$ connections (links) between ROIs in total. The average degree, $<k>$, is $35.80$. The characteristic path length, $L$, is $3.07$ and the clustering coefficient, $C$, is $0.47$. For comparison, the ER networks with the same number of nodes and links yield $L=2.22$ and $C=0.036$ (average over $100$ generated networks). The high small-world index $\sigma_{\text{sw}}$ value of $9.27$, as well as the high $C$ value compared to $C_{\text{rand}}$, suggests that the human brain connectome is a small-world network. On the other hand, it is interesting to note that $L$ is slightly larger than $L_{\text{rand}}$ which suggests the opposite. It is due to the fact that $L$ can be reduced drastically by only a few extremely long-range connections. While the ER networks can have such long-range connections, the connection range of human connectome is relatively limited. The total wiring length $W$ is $493.5$ m. The modularity $Q$ is $0.26$.

For the {\it C. elegans} micro-connectome of $N=279$ neurons and $E=2,287$ links, $L=2.43$ and $C=0.34$ whereas $L_{\text{rand}}=2.30$ and $C_{\text{rand}}=0.059$, which gives the small-world index $\sigma_{\text{sw}}=5.37$. Similar observations can be made as the case of human connectome: $C$ and $\sigma_{\text{sw}}$ indicate that the {\it C. elegans} connectome is a strongly small-world network, but its $L$ is sightly larger than $L_{\text{rand}}$ due to the lack of extremely long-range connections. The total wiring length $W$ is $588.2$ mm. The modularity $Q$ is $0.15$. 

From these basic measures, the connection specificity of the networks can be roughly depicted. Both networks are small-world with few long-range connections and have modular organisation. Since the modularity values, $0.26$ for human and $0.15$ for {\it C. elegans}, are small compared to those of other networks known to have modular structure, the significance of the modular organisation could be questioned. However, these networks, though small, do have modularity indicated by the values when compared to the completely randomized, zero-modularity networks obtained by the link swapping as seen below. 

\begin{table*}
\caption{
{\bf Network measures depicting the connectomes.} 
Network measures for the human brain network with $998$ nodes and the {\it C. elegans} neuronal network with $279$ nodes: $Q$ modularity; $D$ dispersion; $L$ characteristic path length; $C$ clustering coefficient; $\sigma_{\text{sw}}$ small-world index. The values for ER random networks, $L_{\text{rand}}$ and $C_{\text{rand}}$, show the average and the standard deviation over $100$ ER networks.
}
\begin{tabular}{llllllll}
\hline
& $Q$ & $D$ & $L$ & $L_{\text{rand}}$ & $C$ & $C_{\text{rand}}$ & $\sigma_{\text{sw}}$ \\
\hline
Human & $0.26$ & $0.12$ & $3.07$ & $2.231\pm0.001$ & $0.47$ & $0.036\pm0.002 $ & $9.27$ \\
{\it C. elegans} & $0.15$ & $0.46$ & $2.43$ & $2.300\pm0.002$ & $0.34$ & $0.059\pm0.001$ & $5.37$ \\
\hline
\end{tabular}
\begin{flushleft}
\end{flushleft}
\label{table1}
\end{table*}

\begin{figure*}
\includegraphics[width=\textwidth]{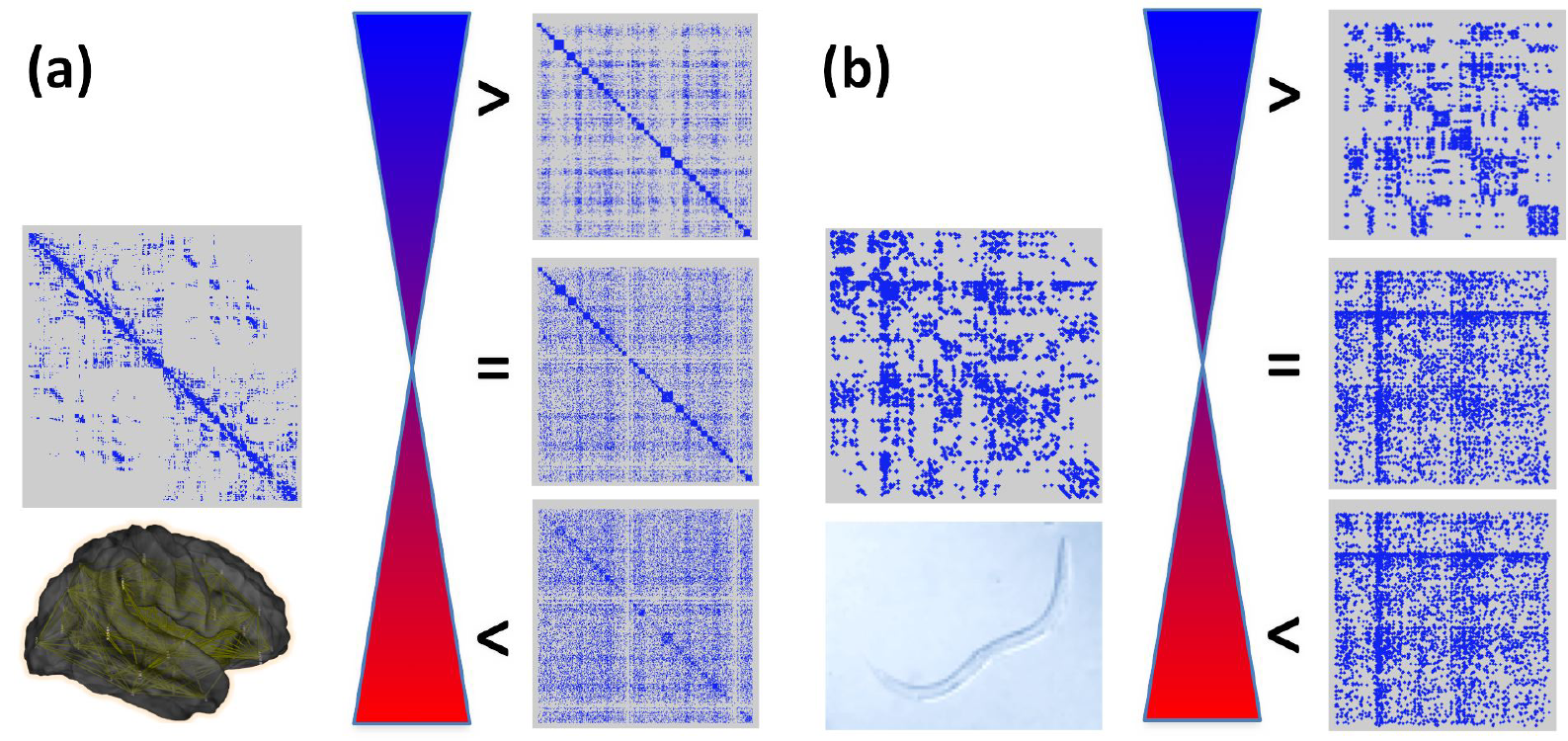}
\caption{
{\bf Adjacency matrices of the connectomes.} The matrices represent the network of ROIs for the human brain (a) and the network of neurons for {\it C. elegans} (b), respectively. Each dot represents a fibre tract between ROIs in (a) or an axonal connection between neurons in (b). For both humans and {\it C. elegans}, we analyzed benchmark networks with similar ($=$), increased ($>$), or decreased ($<$) modularity $Q$ relative to the original neural networks.}
\label{Figure1}
\end{figure*}

To understand the connectivity in detail, next we compared the network measures of the connectome to their respective benchmark networks which were generated through the link-swapping process controlled by simulated annealing as described in Methods. Each node of the benchmark networks has one-to-one correspondence to a node of the original network, and has the same degree and membership to a module as the original node. By changing the control parameter $T$ of the simulated annealing process, the resulting benchmark networks with varying modularities were obtained. The relation between $T$ and resulting $Q$ is given in the Supplementary Figure S2 and Supplementary Table S1. Figure \ref{Figure1} visualises the original neural networks and corresponding benchmark networks with different modularities. 

\begin{figure}
\includegraphics[width=0.5\textwidth]{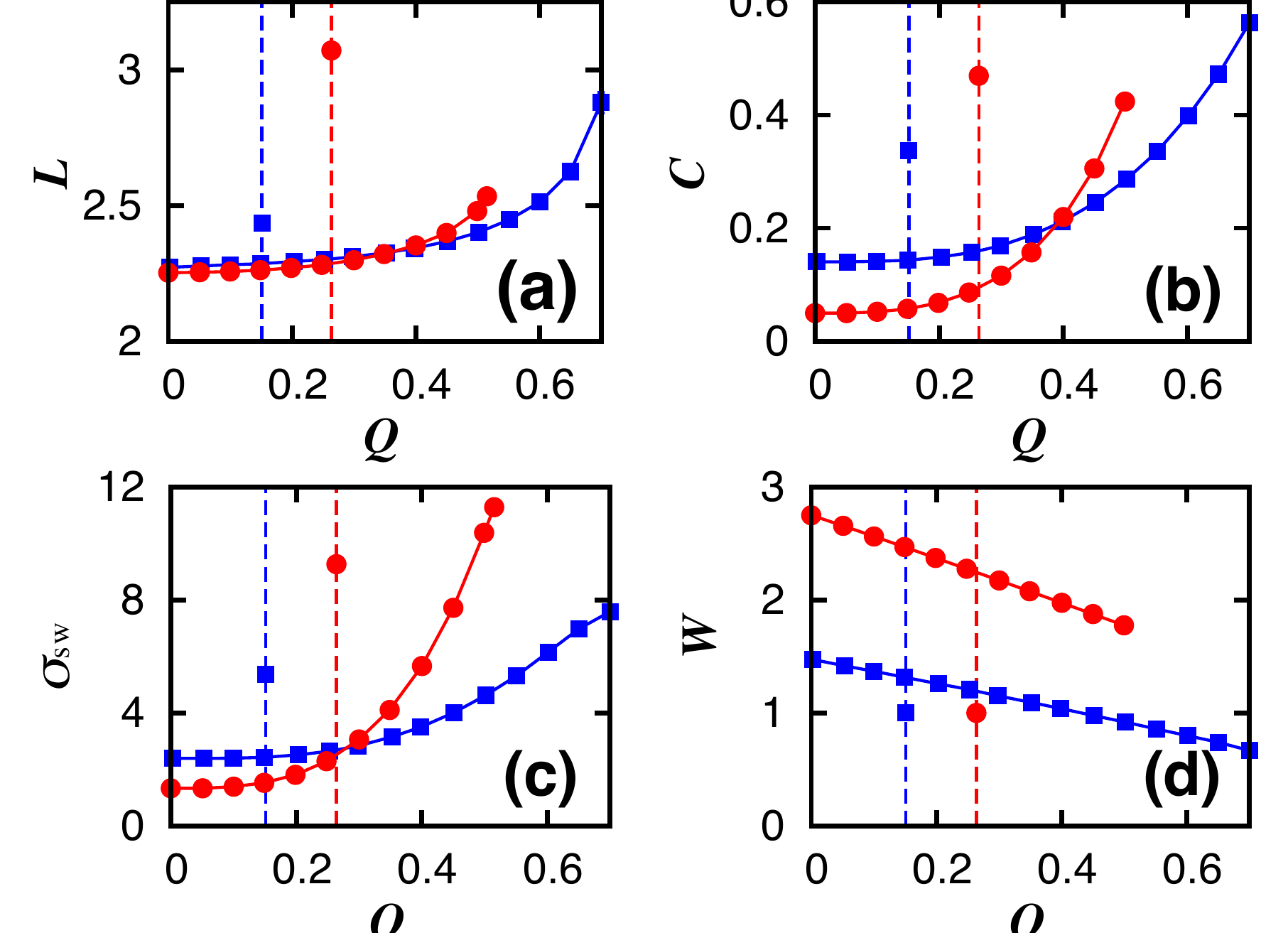}
\caption{
{\bf Small-worldness are different in the connectomes.} The small-world measures, characteristic path length $L$, clustering coefficient $C$, small-world index $\sigma_{\text{sw}}$, and total wiring length $W$, of human (\textcolor{red}{$\bullet$}) and {\it C. elegans} (\textcolor{blue}{$\blacksquare$}) connectomes with respect to modularity, $Q$, which is varied by link swapping. Note that $W$ is normalised with respect to the values of the original neural networks. Unobservable error bars lie within the symbols. The vertical dashed lines denote the values of the original networks. The original networks show more global segregation (higher $L$ suggests lower global efficiency) and more local integration (higher $C$ suggests higher local efficiency) at the same time.}
\label{Figure2}
\end{figure}

The network measures of the original and benchmark networks are shown in Figure \ref{Figure2}. The quantities, $L$, $C$, $\sigma_{\text{sw}}$, and $W$, show strong positive or negative correlations to $Q$ for the benchmark networks, whereas the values from the original network deviate from the trends of the curves in all cases. In general, as the modularity grows, the number of local loops increases and the number of long-range connections decreases. Therefore, the increase in $L$ and $C$, as well as the decrease in $W$, with respect to growing $Q$ is easily understood. For all the network measures, the original neural networks show marked differences to alternative arrangements with the same modularity. In addition, some values for the original networks can only be reached for much higher modularity in alternative networks or cannot be reached at all ($L$, $C$, and $W$ for the human connectome). Note that the clustering coefficients of the original networks are higher than those of alternative networks of the same modularity, which suggests better local interaction efficiency. The high characteristic path length, on the other hand, suggests reduced global communication efficiency. 

\begin{figure*}
\includegraphics[width=\textwidth]{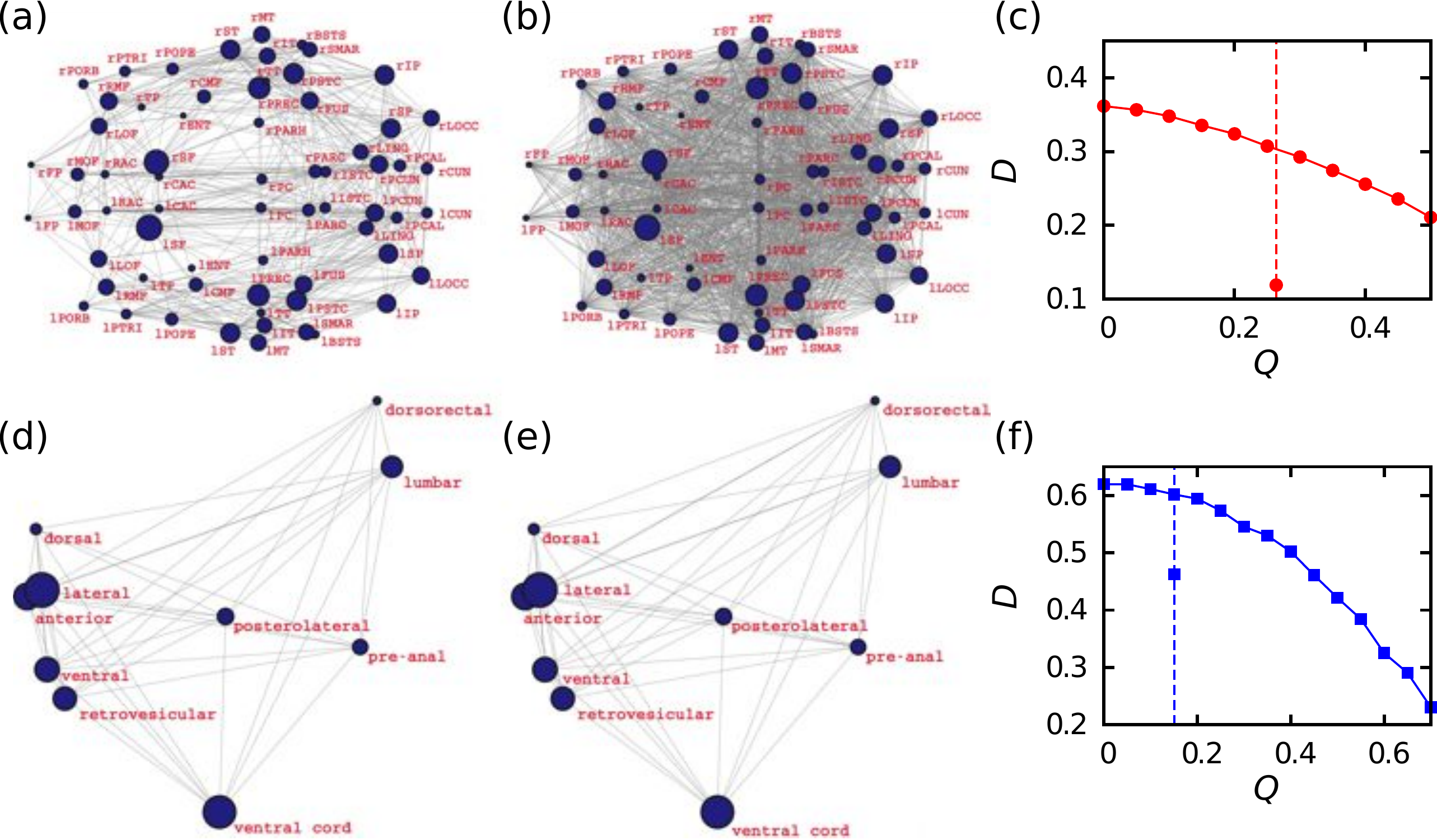}
\caption{
{\bf Connectivity of modules and the dispersion.} The networks of modules for the (a) human (brain areas; horizontal plane) and (d) {\it C. elegans} (ganglia; lateral view) connectomes, and those of the benchmark network snapshots with similar modularity, (b) and (e), respectively. Each node is a module of the networks, whose size is proportional to the square-root of the number of nodes in the module. The locations are given by the centres of mass of its constituent ROIs or neurons. Note that the node locations for {\it C. elegans} are scaled differently in x- and y-axis for visualisation, and do not represent the actual coordinates. The dispersion $D$ of human (c) and {\it C. elegans} (f) networks (data point on the dashed vertical line) is much lower than those of the benchmark networks.}
\label{Figure3}
\end{figure*}

What made the original neural networks deviate from the tendency of alternative benchmark networks, or what is specific to the connectivity of the original networks? The answer is that one module of the neural networks is connected only to a small number of other modules, and corollarily, a pairs of modules are connected to each other by a redundant number of links. A pair of modules are considered to be connected to each other if any member nodes of them are connected. To test such connectivity between modules, the network of modules for both connectomes and examples of their benchmark networks were visualised in Figure \ref{Figure3}. A visual inspection immediately shows that Figure \ref{Figure3}a for the human connectome is sparse and Figure \ref{Figure3}b for a benchmark network of it is dense. This effect is also visible, though less apparent, from Figure \ref{Figure3}d for the {\it C. elegans} connectome and Figure \ref{Figure3}e for the benchmark network. 

As discussed above, the number of links before and after the link swapping does not change. Therefore, the observed difference in link density must have come in during the process of coarse-graining the network of nodes into the network of modules. Note that the multiple number of links between a pair of modules converge into a single link on the network of modules. Accordingly, the number of links on the network of modules is determined by the number of other modules the modules are connected to. Sparse connectivity of the network of modules implies that each module is connected to only a small number of other modules on the network of nodes and that a pair of modules are connected to each other by a redundant number of links. This is observed as bundling of fibres towards relatively few target nodes in the brain connectome, and it is also found in {\it C. elegans} connectome, where neurons are able to follow early established pathways, e.g. in the ventral cord \cite{varier}. On the other hand, the benchmark networks lose such connection specificity during the link swapping. A part of the multiple links from a module to another in the original networks are redirected to multiple number of new modules during the link swapping process, making the number of modules to which they connect larger but the number of links between a given pair of modules smaller.

As a way to measure this, we introduced a novel network property called the dispersion $D$. It measures the average proportion of modules to which a network node is connected. Note, that this is different from an existing measure, the participation coefficient, which is the proportion of a node's connections that connects to other modules, as the dispersion also indicates to \textit{how many} other modules a node is connected to.  For the human connectome, the dispersion is $0.12$, indicating that each ROI is, on average, connected to $12\%$ of all anatomical brain areas (Figure \ref{Figure3}c). For {\it C. elegans} connectome, with a dispersion of $0.46$, each neuron is, on average, connected to $46\%$ of all ganglia (Figure \ref{Figure3}f). These values for the connectomes are much lower than those of the benchmark networks with similar modularity. Human benchmark networks with $Q=0.25$ have $D=0.31$ (larger than the value of human connectome by factor of $2.6$) and {\it C. elegans} benchmark networks with $Q=0.15$ have $D=0.60$ (factor of $1.3$). In addition, such low dispersion values can only be reached for much higher modularities in alternative networks of {\it C. elegans}, or cannot be achieved at all for alternatives of human connectome. Less distributed fibres also reduce the total wiring length, meaning that less energy is needed in connection establishment (myelination) and maintenance (recovery to the resting potential after transmitting an action potential) \cite{laughlin,chklovskii,karbowski}.

These considerations on the costs of material and energy can be seen as related to the physical structure or `hardware' of neural networks. However, costs of the neural `hardware' are not the only potential evolutionary constraint \cite{Kaiser2011Network}. Complementary to the concept of `hardware', the rules for changing the pattern of connections and connection weights can be considered as the `software' of the brain. Connection weights can adapt through learning and connection can be rewired after a lesion or traumatic brain injury \cite{Johansen-Berg2007CurrBiol}. However, looking at changes during brain development, the early perinatal large-scale architecture seems to be remarkably stable. Eliminating activity propagation by blocking neurotransmitter release has little effect on the layer and cortico-cortical architecture \cite{verhage}. Such invariance in the organisation of neural systems could be considered as determined by genetics factors. 

Hence, following question can be raised: how much genetic information is needed to encode the connectivity patterns in human and {\it C. elegans}? One estimate, based on earlier studies in metabolic networks \cite{nykter}, is the algorithmic entropy or Kolmogorov complexity \cite{li}. The algorithmic entropy is the length of a ``sentence'' describing an object in a ``language''. The upper bound of the amount of information embedded in any type of data, here the connectivity matrix, can be approximated by the size of compressed data compared to the size of original data. It can be simply calculated by saving the data in a standard format and then applying a data compression. The compression ratio is the size of the compressed data divided by the size of the original data in bytes. The compression ratio approaches $1$ when almost the same amount of information is needed to describe a network structure, whereas the ratio is close to $0$ when little information is needed to encode the connectivity. In biological terms, we can think of the compressed data as the genetic information, the decompression algorithm as the pattern formation mechanism that is guided through genetic factors, and the uncompressed connectivity matrix as the organisation of neural system that follows neural development.  

\begin{figure}
\includegraphics[width=0.5\textwidth]{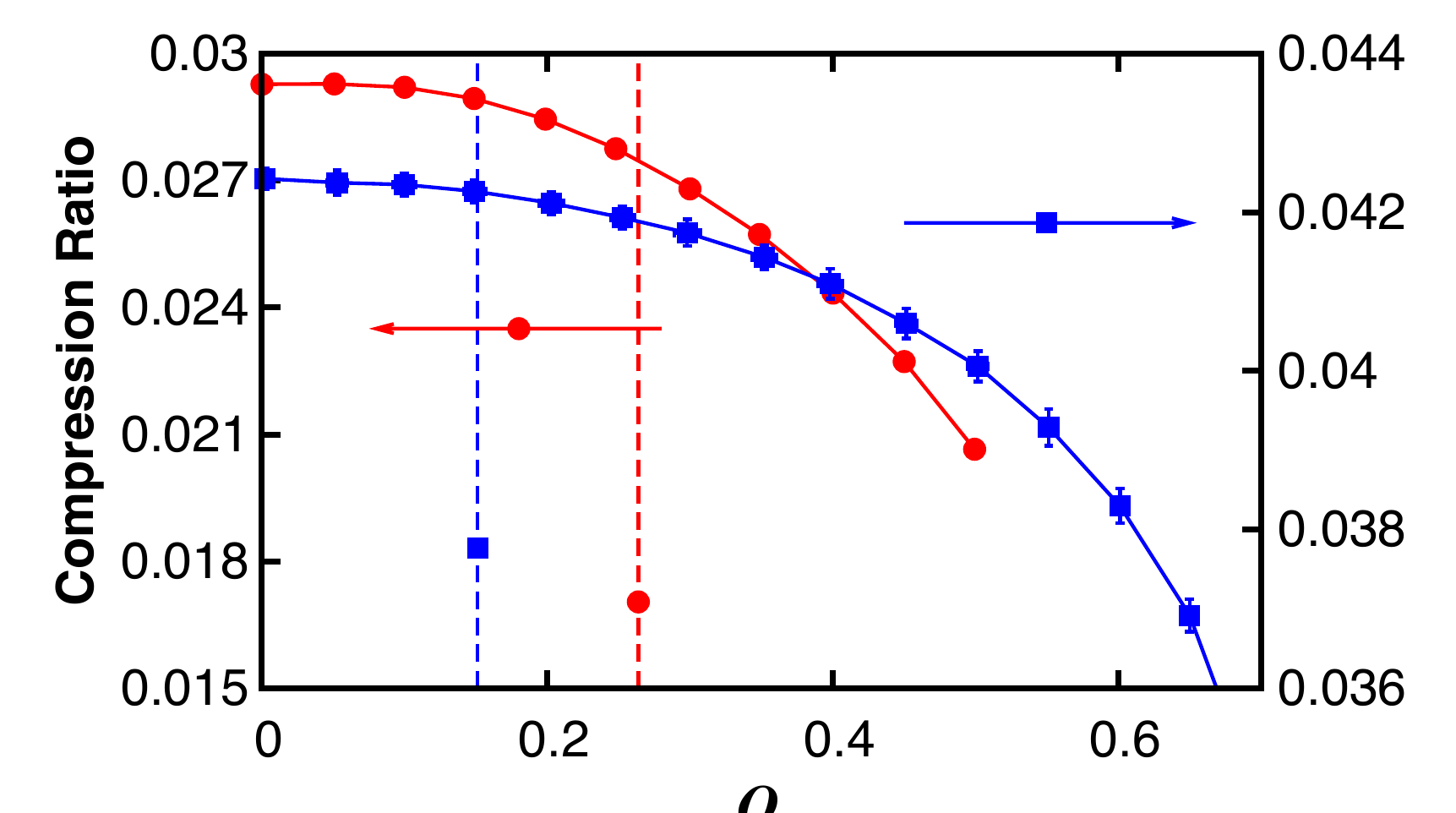}
\caption{
{\bf Compression ratio as a function of modularity.} The compression ratio is defined as the size of the compressed network divided by the size of the original network in bytes when the networks are represented by the adjacency matrices. It is shown for the original (vertical dashed lines) and rewired networks of human (\textcolor{red}{$\bullet$}, left axis) and {\it C. elegans} (\textcolor{blue}{$\blacksquare$}, right axis).
}
\label{Figure4}
\end{figure}

As shown in Figure \ref{Figure4}, the amount of information in the benchmark networks decreases as modularity grows larger. The networks with locally constrained connections are easier to describe than those with many long-range connections, thus have less information. The original networks, however, largely deviate from the curve. The values are comparable to, or even smaller than, the case of maximum modularity. This is also a consequence of the abundant connections between modules. Even when there are a considerable number of connections that are not locally confined, they can be easily described if the connections direct towards similar destinations. The connection specificity of the human connectome, which is locally dense and has only a limited number of global connections between brain areas, requires less information in describing the topology. Similar observations and arguments are applied to the {\it C. elegans} connectome as well.

\section{Discussion}
Neural systems show a modular architecture at different hierarchical levels, ranging from the network of individual neurons to network of brain regions. Observing human and {\it C. elegans} neural networks, we showed that the original networks are markedly different from the alternative benchmark networks. From both of the connectomes, we found the evidences indicate that local information distribution is more efficient but global integration is less so by studying the clustering coefficient and the characteristic path length, respectively. We also found that metabolic costs for establishing neural connections are low, which is suggested by relatively small total wiring length. To explain these results with the connection specificity of the neural networks, we introduced the novel measure dispersion, the ratio of modules to which an individual node is connected on average. By quantifying the distribution of connections across the modules, we found that smaller dispersion is specific to the original neural networks. Third, both neural networks showed a low algorithmic entropy, which indicates less requirement for the rules to organise the architecture of neural networks. 

\subsection*{Increased separation reduces spreading and interference}
Characteristic path lengths of neural networks were high compared to benchmark networks with the same modularity. Relatively high path length makes rapid spreading of activity less likely, as for epileptic seizures \cite{kaiser_lsa}. Sparse connectivity between modules can become a bottleneck for information flow. On the other hand, higher connectivity between modules or merging of modules can enhance the likelihood of activity propagation---we previously described this bottleneck behaviour as {\it topological inhibition} \cite{kaiser_crit}.  Recent studies of functional connectivity in epilepsy indeed found a reduced path length, measured through increased global efficiency, and more connections between modules  \cite{chavez}. Therefore, a relatively large characteristic path length might be one of the features that support healthy cognitive functions \cite{kaiser_nonoptimal,chen,vertes}.  At the same time, increased neighbourhood connectivity, as measured by high clustering coefficient, renders a strong local interaction possible within a functionally related brain area or ganglion. In the similar line of argument, a study of oscillatory dynamics on neural networks has shown that the modular structure enables strong synchronization within modules and weak between them \cite{zhao}.

\subsection*{Reduced dispersion decreases total wiring length}
Low total wiring length reduces metabolic costs for connection establishment and at the same time obstacles activity propagation in neural systems \cite{kaiser_nonoptimal,chen,cherniak}. For both the human and {\it C. elegans} connectome, we saw reduced dispersion of connections which is linked to the decreased total wiring length. As primate and nematode systems are close to the optimal arrangement for reducing wiring length \cite{kaiser_nonoptimal,Markov2012CerCor}, any re-arrangement of connections to spread more widely throughout the network will lead to the formation of longer connections in the system. A mechanism that can limit dispersion in fibre tract systems is fasciculation of axons. The fasciculation is a mechanism that a small number of pioneer neurons form pathways that guide the axons of the following neurons, resulting in a bundle of axon fibres. This might also be the case for {\it C. elegans} where some neurons in the ventral cord are formed early on \cite{varier} providing a pathway between anterior and posterior parts of the worm. The reliance on pioneer fibres might prevent more diverse connectivity to other areas located afar. 

Given the relation between dispersion and other network properties that change in schizophrenia \cite{zalesky}, autism \cite{courchesne}, or epilepsy \cite{walker}, a reduced coherence of fibre tracts might be an important component in the path towards developmental diseases. Moreover, the dispersion might be related to changes in diffusion imaging, since a more distributed pattern of connectivity would break apart the fascicular pattern of fibre tracts. Therefore, we would expect that higher values of dispersion are associated with lower values of fractional anisotropy (FA) and to a shift towards more regular networks with higher characteristic path length as well as clustering coefficient. For neural disorders, for example, a shift towards regular networks has been reported for epilepsy \cite{ponten} and lowered FA was reported for schizophrenia \cite{skudlarski,heuvel}. Note, however, that lower FA might not only result from more diffuse fibre tracts within a voxel but also from reduced myelination.   

\subsection*{Development of modular neural networks}
Both of the connectomes showed higher Kolmogorov complexity as measured through the compression ratio. This algorithmic entropy is different from the information theory inspired entropy, which has been applied to brain networks \cite{sporns}. Kolmogorov complexity shows how much code is needed to generate an object. The generation of neural networks is the process of neural development. It can be driven by several factors including genes, epigenetic factors, and self-organisation. Although we only begin to understand the relation between genes and connectomes \cite{kaufman,fornito}, it has been pointed out that gene expression patterns which mediate growth factors and guidance cues play an important role in determining the connectivity of neural systems \cite{ooyen}. However, gene expression and the inclusion of genes into the genome are costly endeavours that would be expected to be under evolutionary pressure. Indeed, neural systems try to reduce the amount of genetic encoding that is needed for neural networks. At early stages of development in {\it C. elegans}, most long-distance connections can be established when the neurons are nearby \cite{varier}. This can reduce the need to control axon growth over long distances. The lower dispersion, which we found in both connectomes, might be another mechanism to reduce the amount of code requirement. Altogether, this suggests that the neural system might be efficient not only for the metabolic `running costs' \cite{laughlin} but also in terms of their developmental mechanisms. 

Which developmental mechanism could influence the modular organisation of neural systems? Several potential biological mechanisms for generating hierarchical modular networks have been described. One way is to start with an existing network and generate copies of the network where the copies retain the same internal connectivity as the original network but also establish connections directly to the original network. Variations of this method can be used to generate hierarchical scale-free networks \cite{ravasz} and were also thought to lead to cortical connectivity-like networks \cite{krubitzer}. The timing of synaptogenesis and cell birth can also be crucial for development \cite{varier,rakic}. For modular networks, time windows during development can lead to multiple modules where the module number, module size, and inter-module-connectivity is determined by the number, width and overlap of developmental time windows for synaptogenesis, respectively \cite{nisbach, kaiser_growth}. 

\subsection*{Link swapping perturbs lattice structure}
Neural systems can be seen as lattice networks, using two-dimensional sheets of tissue preferring to connect to nearby nodes \cite{kaiser_rule} with additional long-distance shortcuts to promote rapid processing and integration of information \cite{kaiser_nonoptimal}. The connections are established with geometrical constraints \cite{henderson}. Previous studies have shown that lattice networks show a low compression ratio compared to other topologies \cite{sun}. During the link swapping, however, such geometrical constraints becomes relaxed. A rewired link can establish a new connection with any node in the module (intra-module link) or any node in the entire network (inter-module link). As a result, the lattice structure of the original network is perturbed which leads to higher dispersion. High dispersion prohibits efficient compression and the Kolmogorov complexity of the perturbed networks becomes high. It has been claimed that other measures of the neural networks, such as characteristic path length, clustering coefficient, and modularity, can also be interpreted as those of regular networks \cite{henderson}. The current study rediscovers such findings by showing that perturbation in lattice structure makes those measures deviate from the original values. 

\subsection*{Conclusions}
In summary, both {\it C. elegans} and the human connectome show reduced global efficiency (higher characteristic path length), increased local efficiency (higher clustering coefficient), and reduced metabolic cost (lower total wiring length) compared to random modular networks. A marked difference in the organisation of the connectomes that is relevant to those properties is their low dispersion. The specific modular organisation of the connectomes requires fewer rules to construct it (lower algorithmic entropy), or fewer genetic factors to develop such neural system. Together, these results show that neural systems across different levels, from the network of neurons to the network of brain regions, commonly have efficiencies in multiple aspects listed above. The hierarchical natures of the modular organisation of these connectomes and how they can be understood with respect to the multiple constraints given by various network measures \cite{meunier, krumnack} remain a topic for future studies.

\ack{The authors are grateful to Olaf Sporns for sharing the connectome data. This research was supported by the WCU program through the KOSEF funded by the MEST (R31-10089), EPSRC (EP/K026992/1), and the CARMEN e-science project (\protect{http://www.carmen.org.uk}) funded by EPSRC (EP/E002331/1).}

% \bibliographystyle{plain}
% \bibliography{modularity_ref}

\begin{thebibliography}{99}

\bibitem{kaiser_tutorial}
[1] Kaiser M. 2011 A tutorial in connectome analysis: Topological and spatial features of brain networks. {\em Neuroimage}, {\bf 57}, 892--907.

\bibitem{mackie}
[2] Mackie GO. 1988 Nerve nets. {\em Comparative neuroscience and neurobiology}, 84--86. Springer, New York.

\bibitem{arendt}
[3] Arendt D, Denes AS, Jekely G, Tessmar-Raible K. 2008 The evolution of nervous system centralization. {\em Phil Trans R Soc Lond B Biol Sci}, {\bf 363}, 1523--1528. 

\bibitem{white}
[4] White JG, Southgate E, Thomson JN, Brenner S. 1986 The structure of the nervous system of the nematode \textit{{C}.
  elegans}. {\em Phil Trans R Soc Lond B Biol Sci}, {\bf 314}, 1--340.

\bibitem{kaiser_dyn}
[5] Kaiser M, Hilgetag CC, K{\"{o}}tter R. 2010 Hierarchy and dynamics of neural networks. {\em Front Neuroinformatics}, {\bf 4}, 112.

\bibitem{meunier}
[6] Meunier D, Lambiotte R, Bullmore ET. 2010 Modular and hierarchically modular organization of brain networks. {\em Front Neurosci}, {\bf 4}, 200.

\bibitem{krumnack}
[7] Krumnack A, Reid AT, Wanke E, Bezgin G, K{\"o}tter R. 2010 Criteria for optimizing cortical hierarchies with continuous ranges. {\em Front Neuroinfom}, {\bf 4}, 7. 

\bibitem{newman}
[8] Newman MEJ. 2004 Fast algorithm for detecting community structure in networks. {\em Phys Rev E Stat Nonlin Soft Matter Phys}, {\bf 69}, 066133.

\bibitem{Lim2014CerCor}
[9] Lim S, Han C, Uhlhaas PJ, Kaiser M. 2013 Preferential detachment during human brain development: Age- and sex-specific structural connectivity in {D}iffusion {T}ensor {I}maging ({DTI}). {\em Cereb Cortex}, Advanced online.

\bibitem{seung}
[10] Seung HS. 2009 Reading the book of memory: sparse sampling versus dense mapping of connectomes. {\em Neuron}, {\bf 62}, 17--29.

\bibitem{defelipe}
[11] DeFelipe J. 2010 From the connectome to the synaptome: An epic love story. {\em Science}, {\bf 330}, 1198--1201.

\bibitem{sritharan}
[12] Sritharan S, Han CE, Rotarska-Jagiela A, Singer W, Deichmann R, Maurer K, Kaiser M, Uhlhaas PJ. Increased long-range connectivity in schizophrenia: A Graph theoretical analysis of Diffusion Tensor Image data. (submitted) 

\bibitem{chavez}
[13] Chavez M, Valencia M, Navarro V, Latora V, Martinerie J. 2010 Functional modularity of background activities in normal and epileptic brain networks. {\em Phys Rev Lett}, {\bf 104}, 118701.

\bibitem{roberts}
[14] Roberts A, Conte D, Hull M, Merrison-Hort R, al~Azad AK, Buhl E, Borisyuk R, Soffe SR. 2014 Can Simple Rules Control Development of a Pioneer Vertebrate Neuronal Network Generating Behavior? {\em J Neurosci}, {\bf 34}, 608--621.

\bibitem{hagmann}
[15] Hagmann P, Cammoun L, Gigandet X, Meuli R, Honey CJ, {\it et al.} 2008 Mapping the structural core of human cerebral cortex. {\em PLoS Biol}, {\bf 6}, e159.

\bibitem{varier}
[16] Varier S, Kaiser M. 2011 Neural development features: Spatio-temporal development of the \textit{{C}. elegans} neuronal network. {\em PLoS Comput Biol}, {\bf 7}, e1001044.

\bibitem{ay}
[17] Achacoso TB, Yamamoto WS. 1992 {\em AY's Neuroanatomy of \textit{{C}. elegans} for Computation}. CRC Press, Boca Raton, FL.

\bibitem{watts}
[18] Watts DJ, Strogatz SH. 1998 Collective dynamics of `small-world' networks. {\em Nature}, {\bf 393}, 440--442.

\bibitem{humphries}
[19] Humphries M, Gurney K. 2008 Network `small-worldness': a quantitative method for determining canonical network quivalence. {\em PLoS ONE}, {\bf 3}, e0002051.

\bibitem{guimera}
[20] Guimera R, Amaral LA. 2005 Cartography of complex networks: modules and universal roles. {\em J Stat Mech}, 2005(P02001):nihpa35573.

\bibitem{li}
[21] Li M, Vitanyi P. 1997 {\em An Introduction to Kolmogorov Complexity and Its Applications}. Springer, New York, second edition.

\bibitem{ziv}
[22] Ziv J, Lempel A. 1977 Universal algorithm for sequential data compression. {\em IEEE T Inform Theory}, {\bf 23}, 337--343.

\bibitem{rubinov}
[23] Rubinov M, Sporns O. 2010 Complex network measures of brain connectivity: Uses and interpretations. {\em Neuroimage}, {\bf 52}, 1059--1069.

\bibitem{laughlin}
[24] Laughlin SB, de~Ruyter~Van~Steveninck RR, Anderson JC. 1998 The metabolic cost of neural information. {\em Nat Neurosci}, {\bf 1}, 36--41.

\bibitem{chklovskii}
[25] Chklovskii DB, Koulakov AA. 2004 Maps In The Brain: What Can We Learn from Them? {\em Annu Rev Neurosci}, {\bf 27}, 369--392.

\bibitem{karbowski}
[26] Karbowski J. 2014 Constancy and trade-offs in the neuroanatomical and metabolic design of the cerebral cortex. {\em Front Neural Circuits}, {\bf 8}, 9.

\bibitem{Kaiser2011Network}
[27] Kaiser M, Varier S. 2011 Evolution and development of brain networks: From \textit{{C}aenorhabditis elegans} to \textit{{H}omo sapiens}. {\em Network: Computation in Neural Systems}, {\bf 22}, 143--147.

\bibitem{Johansen-Berg2007CurrBiol}
[28] Johansen-Berg H. 2007 Structural plasticity: rewiring the brain. {\em Curr Biol}, {\bf 17}, R141--4.

\bibitem{verhage}
[29] Verhage M, Maia A, Plomp JJ, Brussaard AB, Heeroma JH, {\it et al.} 2000 Synaptic assembly of the brain in the absence of neurotransmitter secretion. {\em Science}, {\bf 287}, 864--869.

\bibitem{nykter}
[30] Nykter M, Price ND, Larjo A, Aho T, Kauffman SA, {\it et al.} 2008 Critical networks exhibit maximal information diversity in structure-dynamics relationships. {\em Phys Rev Lett}, {\bf 100}, 058702.

\bibitem{kaiser_lsa}
[31] Kaiser M, Hilgetag CC. 2010 Optimal hierarchical modular topologies for producing limited sustained activation of neural networks. {\em Front Neuroinformatics}, {\bf 4}, 8.

\bibitem{kaiser_crit}
[32] Kaiser M, G{\"{o}}rner M, Hilgetag CC. 2007 Functional criticality in clustered networks without inhibition. {\em New J Phys}, {\bf 9}, 110.

\bibitem{kaiser_nonoptimal}
[33] Kaiser M, Hilgetag CC. 2006 Nonoptimal component placement, but short processing paths, due to long-distance projections in neural systems. {\em PLoS Comput Bio}, {\bf 2}, e95.

\bibitem{chen}
[34] Chen Y, Wang S, Hilgetag CC, Zhou C. 2013 Trade-off between Multiple Constraints Enables Simultaneous Formation of Modules and Hubs in Neural Systems. {\em PLoS Comput Biol}, {\bf 9}, e1002937.

\bibitem{vertes}
[35] V{\'e}rtes PE, Alexander-Bloch AF, Gogtay N, Giedd JN, Rapoport JL, Bullmore ET. 2012 Simple models of human brain functional networks. {\em Proc Natl Acad Sci USA}, {\bf 109}, 5868--5873.

\bibitem{zhao}
[36] Zhao M, Zhou C, L{\"u} J, Lai CH. 2011 Competition between intra-community and inter-community synchronization and relevance in brain cortical networks. {\em Phy Rev E}, {\bf 84}, 016109. 

\bibitem{cherniak}
[37] Cherniak C. 1994 Component placement optimization in the brain. {\em J Neurosci}, {\bf 14}, 2418--2427.

\bibitem{Markov2012CerCor}
[38] Markov NT, Ercsey-Ravasz MM, Ribeiro~Gomes AR, Lamy C, Magrou L, Vezoli J, Misery P, Falchier A, Quilodran R, Gariel MA, Sallet J, Gamanut R, Huissoud C, Clavagnier S, Giroud P, Sappey-Marinier D, Barone P, Dehay C, Toroczkai Z, Knoblauch K, Van~Essen DC, Kennedy H. 2014 A weighted and directed interareal connectivity matrix for macaque cerebral cortex. {\em Cereb Cortex}, {\bf 24}, 17--36.

\bibitem{zalesky}
[39] Zalesky A, Fornito A, Seal ML, Cocchi L, Westin CF, {\it et al.} 2011 Disrupted axonal fiber connectivity in schizophrenia. {\em Biol Psychiatry}, {\bf 69}, 80--89.

\bibitem{courchesne}
[40] Courchesne E, Pierce K. 2005 Why the frontal cortex in autism might be talking only to itself: local over-connectivity but long-distance disconnection. {\em Curr Opin Neurobiol}, {\bf 15}, 225--230.

\bibitem{walker}
[41] Walker MC, Daunizeau J, Lemieux L. 2011 Concepts of connectivity and human epileptic activity. {\em Front Syst Neurosci}, {\bf 5}, 12.

\bibitem{ponten}
[42] Ponten S, Bartolomei F, Stam C. 2007 Small-world networks and epilepsy: Graph theoretical analysis of intracerebrally recorded mesial temporal lobe seizures. {\em Clin Neurophysiol}, {\bf 118}, 918--927.

\bibitem{skudlarski}
[43] Skudlarski P, Jagannathan K, Anderson K, Stevens MC, Calhoun VD, {\it et al.} 2010 Brain connectivity is not only lower but different in schizophrenia: A combined anatomical and functional approach. {\em Biol Psychiatry}, {\bf 68}, 61--69.

\bibitem{heuvel}
[44] van~den~Heuvel MP, Mandl RCW, Stam CJ, Kahn RS, Pol HEH. 2010 Aberrant frontal and temporal complex network structure in schizophrenia: A graph theoretical analysis. {\em J Neurosci}, {\bf 30}, 15915--15926.

\bibitem{sporns}
[45] Sporns O. 2006 Small-world connectivity, motif composition, and complexity of fractal neuronal connections. {\em Biosystems}, {\bf 85}, 55--64.

\bibitem{kaufman}
[46] Kaufman A, Dror G, Meilijson I, Ruppin E. 2006 Gene expression of \textit{{C}. elegans} neurons carries information on their synaptic connectivity. {\em PLoS Comput Biol}, {\bf 2}, e167.

\bibitem{fornito}
[47] Fornito A, Zalesky A, Bassett DS, Meunier D, Ellison-Wright I, {\it et al.} 2011 Genetic influences on cost-efficient organization of human cortical functional networks. {\em J Neurosci}, {\bf 31}, 3261--3270.

\bibitem{ooyen}
[48] van~Ooyen A, Willshaw DJ. 2000 Development of nerve connections under the control of neurotrophic factors: parallels with consumer-resource systems in population biology. {\em J Theor Biol}, {\bf 206}, 195--210.

\bibitem{ravasz}
[49] Ravasz E, Somera AL, Mongru DA, Oltvai ZN, Barab{\'a}si AL. 2002 Hierarchical organization of modularity in metabolic networks. {\em Science}, {\bf 297}, 1551--1555.

\bibitem{krubitzer}
[50] Krubitzer L, Kahn DM. 2003 Nature versus nurture revisited: An old idea with a new twist. {\em Prog Neurobiol}, {\bf 70}, 33--52.

\bibitem{rakic}
[51] Rakic P. 2002 Neurogenesis in adult primate neocortex: An evaluation of the evidence. {\em Nat Rev Neurosci}, {\bf 3}, 65--71.

\bibitem{nisbach}
[52] Nisbach F, Kaiser M. 2007 Developmental time windows for spatial growth generate multiple-cluster small-world networks. {\em Euro Phys J B}, {\bf 58}, 185--191.

\bibitem{kaiser_growth}
[53] Kaiser M, Hilgetag CC. 2007 Development of multi-cluster cortical networks by time windows for spatial growth. {\em Neurocomput}, {\bf 70}, 1829--1832.

\bibitem{kaiser_rule}
[54] Kaiser M, Hilgetag CC, van Ooyen A. 2009, A simple rule for axon outgrowth and synaptic competition generates realistic connection lengths and filling fractions. {\em Cereb Cortex}, {\bf 19}, 3001--3010.

\bibitem{henderson}
[55] Henderson JA, Robinson PA. 2011 Geometric Effects on Complex Network Structure in the Cortex. {\em Phys Rev Lett}, {\bf 107}, 018102. 

\bibitem{sun}
[56] Sun J, Bollt EM, ben~Avraham D. 2008 Graph compression---save information by exploiting redundancy. {\em J Stat Mech Theor Exp}, {\bf 06}, P06001.

%\bibitem{costa}
%Costa LdF, Rodrigues FA, Hilgetag CC, Kaiser M. 2009 Beyond the average: detecting global singular nodes from local features in complex networks. {\em Europhys Lett}, {\bf 87}, 18008.

\end{thebibliography}

\end{document}